\documentclass{JINST}

\RequirePackage{xspace}

\def\gev    {\ensuremath{\mathrm{\,Ge\kern -0.1em V}}\xspace}
\def\mev    {\ensuremath{\mathrm{\,Me\kern -0.1em V}}\xspace}

\def\lint   {\ensuremath{\,\lambda}\xspace}

\def\cma    {\ensuremath{{\rm \,cm}^2}\xspace}

\def\mm     {\ensuremath{{\rm \,mm}}\xspace}

\title{Impact of dead zones on the response of a hadron calorimeter with projective and non-projective
geometry}

\author{J. Blaha\thanks{Corresponding author.}, N. Geffroy, and Y. Karyotakis\\
\llap{$ $}Laboratoire d'Annecy-le-Vieux de Physique des Particules, 
Universit\'{e} de Savoie, CNRS/IN2P3,\\
9 Chemin de Bellevue 74980 Annecy-le-Vieux, France\\
E-mail: \email{jan.blaha@lapp.in2p3.fr}}

\abstract{The aim of this study is to find an optimal mechanical design of the hadronic calorimeter for SiD 
detector which takes into account engineering as well as physics requirements. The study focuses on the crack 
effects between two modules for various barrel mechanical design on calorimeter response. The impact
of different size of the supporting stringers and dead areas in an active calorimeter layer along the module 
boundary has been studied for single pions and muons. The emphasis has been put on the comparison of the 
projective and non-projective barrel geometry for SiD hadronic calorimeter.}


\begin{document}

\section{Introduction}
Design of the future particle physics detectors for the International Linear Collider (ILC)~\cite{ilc} is 
optimized for usage of the Particle Flow Algorithm (PFA)~\cite{pfa}. By this strategy, a jet energy resolution 
of $30\%/\sqrt{E}$ can be reached and thus allows the reconstruction of the invariant masses of W's, Z's, and 
tops with resolutions close to the natural widths of these particles. On the other hand, new challenges are put 
on the construction of the detector subsystems, particularly on calorimeters which must have imaging capability 
allowing the assignment of energy cluster deposits to charged or neutral particles with high accuracy. In order 
to fulfill this requirement, several detector technology for active part of the calorimeters are currently under 
development. In parallel, the work on the engineering design of whole calorimeters and their assembly methods are 
well advanced.  
   
The aim of this study is to find an optimal mechanical design of the hadronic calorimeter (HCAL) for SiD 
detector~\cite{sid} which takes into account engineering as well as physics constrains. Therefore, the impact of the 
various HCAL mechanical design on the calorimeter response has been evaluated. Since the calorimeter is composed 
from independent modules, which due to mechanical reasons creates discontinuities in the detection, the study 
is focused on the hadronic shower behavior close to the boundaries between these modules for different calorimeter 
geometry and for size of dead areas along this boundary. 

\section{Calorimeter geometry and simulation tools}
\subsection{Projective and non-projective geometry}
The SiD hadronic calorimeter is a sampling calorimeter, which is located inside the magnet coil and surrounds the 
electromagnetic calorimeter. The barrel part is divided into twelve azimutals modules, each one consisting of 40 layers
with a passive part composed of 1.89~cm thick stainless steel absorber and an active part equipped with RPC chambers 
(SiD baseline detector choice for HCAL) or alternative detectors, such as Micromegas, GEM or 
Scintillator~\cite{sid}. Thus the total calorimeter depth is 4.5~$\lambda$ and the overall dimensions are: 
$l=6036$~mm, $R_{int}=1419$~mm, and $R_{ext}=2583$~mm, where $l, R_{int}, R_{ext}$ denote calorimeter length, and its 
internal and external radii, respectively. 

\begin{center}         
  \begin{figure}[htb]
    \includegraphics[width=0.49\columnwidth]{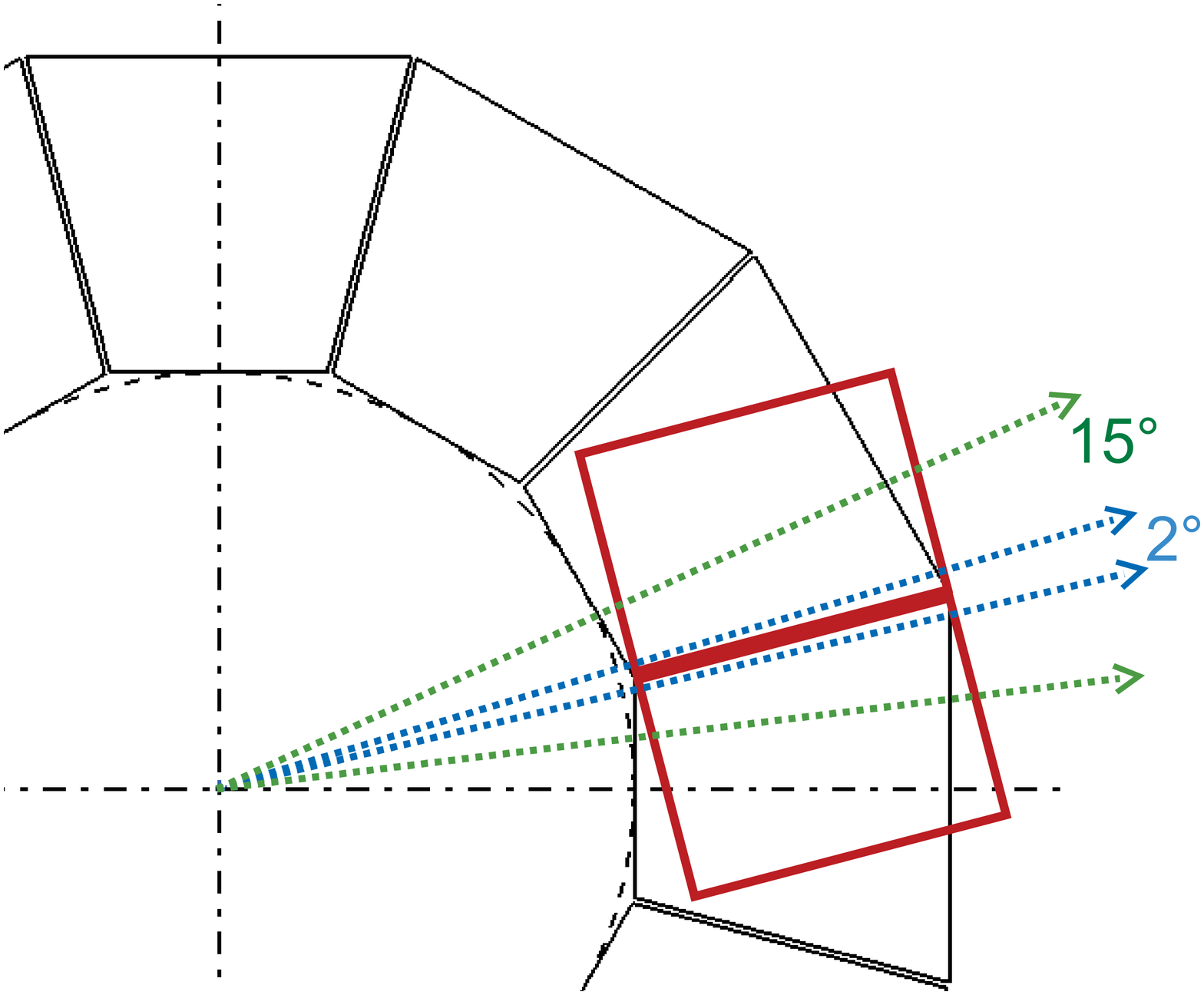}
    \hfill
    \includegraphics[width=0.49\columnwidth]{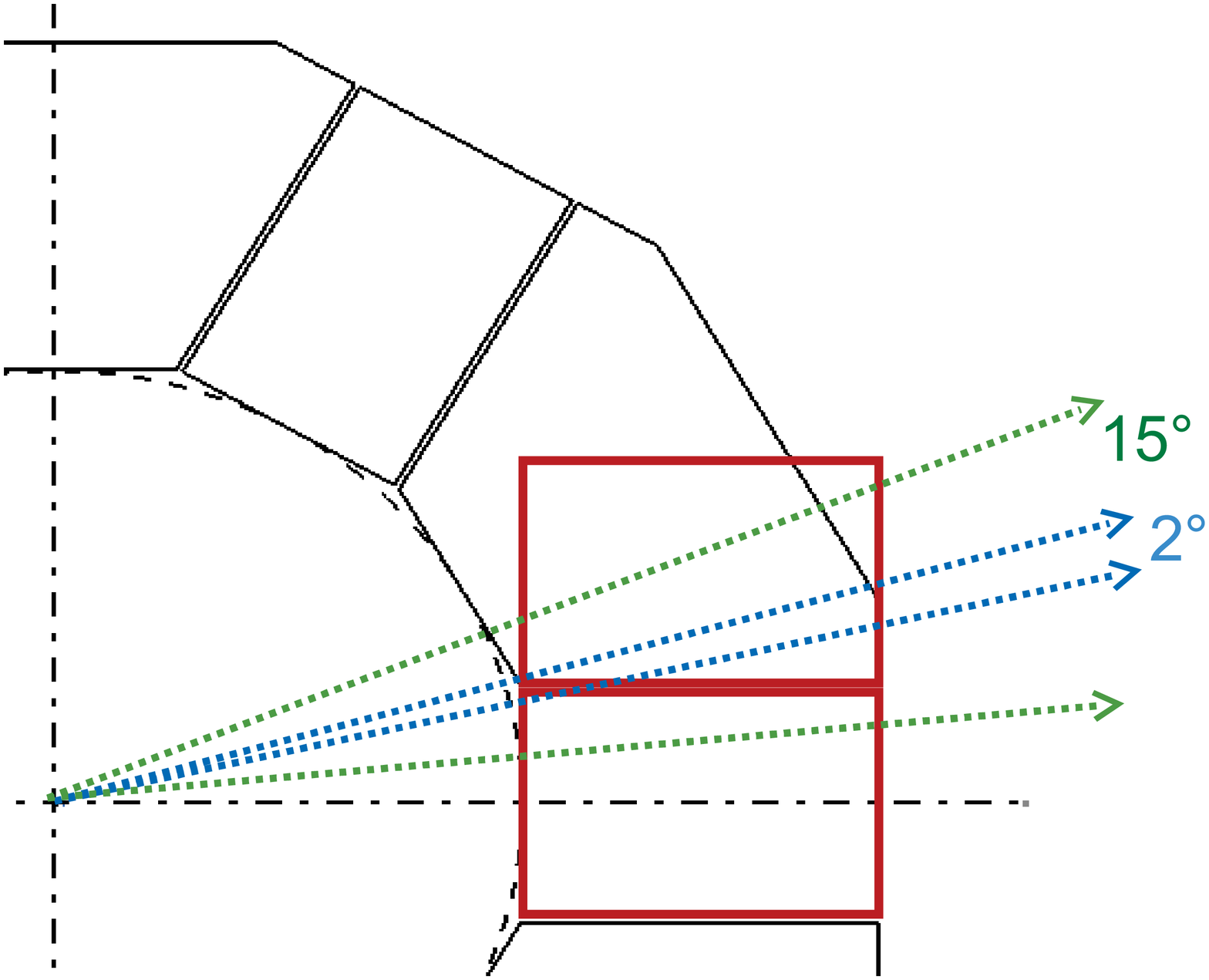}\\ 
    \caption{A quarter of the projective (left) and non-projective (right) calorimeter geometry, both displayed in 
      black color. The simulation configuration of two modules and their position with respect to the interaction 
      point is shown in red color. Blue and green arrows represent cones with a vertex angle of 2$^{\circ}$ or 
      15$^{\circ}$, respectively. The impinging particles are randomly generated within these vertex angles.}
    \label{geometry}
  \end{figure}
\end{center}

Two calorimeter geometries have been proposed so far. The first one is a projective geometry consisting of twelve 
identical trapezoidal modules whose edges are pointing to the beam axis, see Fig.~\ref{geometry} left. The second 
geometry was designed in order to avoid cracks in calorimeter between the modules. Therefore it is an off-pointing 
or non-projective geometry, where the modules boundaries are not projective with respect to the beam axis. It 
consists of six trapezoidal and six rectangular modules, which are arranged on a rota basis as it is shown in 
Fig.~\ref{geometry} right. The shape of the trapezoidal modules of the non-projective geometry is such, that internal 
and external surface of both proposed geometries are identical, so called do-decagon. A detail 
description of the geometry for SiD HCAL can be find in Ref.~\cite{geometry}.   
    
\subsection{Simulation set-up}
\subsubsection{Geometry configurations}
Since the description of the barrel calorimeter (including 12 modules) with all necessary mechanical details
(different shape of the modules and utilization of the supporting stringers) had not been available for a detailed 
simulation study, a simplified geometry using only two adjacent rectangular modules has been implemented. These 
modules are placed under two different angles with respect to the interaction point, as it is illustrated in 
Fig.~\ref{geometry}, which correspond to the boundary position of the projective and non-projective module 
configuration, respectively. Because the study is focused on the boundary effects with single impinging 
particles, the simplified geometry with one module transition can be considered as a good approximation of the 
two modules with trapezoidal shape as well. Moreover, due to the axial symmetry of the calorimeter, the results 
obtained with the simplified geometry can be extended to the whole calorimeter having twelve modules.       

The overall dimensions of each rectangular module is $2000\times2000\times1076$~mm$^3$, where a calorimeter depth 
of 1076~mm is corresponding to 4.5\lint. Each module consists of 40 absorbers made from a 1.89~cm thick stainless steel 
plates interlayed by 8~mm gaps for Micromegas chambers which were chosen as an calorimeter active medium for this study. 
An active part of the chamber is 6.5~mm thick and includes: 1.2~mm of PCB material, 2.3~mm of material for the chips 
and other passive components and 3 mm of gas (Ar/Isobutane 95/5). A detail geometry description of the existing Micromegas 
prototype which has been implemented in the simulation can be found in Ref.~\cite{micromegas}.

Modules in the SiD calorimeter are hold together by supporting stringers made of 2~cm thick stainless steel plates 
which contribute to the dead areas in the calorimeter. In order to evaluate how they affect the calorimeter 
response, configurations with different stringer thickness were considered for the projective and non-projective 
geometry, respectively. The first configuration is ideal with two rectangular modules interconnected without supporting 
plates and hence without any cracks between them. The second and third configurations have 1 and 2~cm thick supporting 
plates, respectively. The last configuration is similar to the third one, but material of the electromagnetic calorimeter, 
which is equal to about 1~$\lambda$, was added in front of the hadronic calorimeter. In this configuration the 
electromagnetic calorimeter is used only as a passive detector component in order to estimate the fraction of the 
hadronic shower deposition within its material. 

\subsubsection{Generated data samples}
Two sets of Monte Carlo data were generated by the GEANT4-based simulator SLIC~\cite{slic} with the
\verb,QGSP_BERT, physics list for the four above described configurations and for the projective and non-projective 
geometry. For each set, 50~GeV negative pions and muons were generated randomly within a vertex cone 
angle of about 2$^{\circ}$ and 15$^{\circ}$, directed from the interaction point toward the modules boundary, see 
Fig.~\ref{geometry}. Therefore an impact area around the boundary was restricted to a disc with a radius 
of 2.5~cm for smaller the cone angle and about 19.1~cm for the larger cone angle. Precise values of the cone angles 
and impact areas for projective and non-projective geometry are displayed in Fig.~\ref{appxGeometry}~(Appx.~A).
The small impact area was studied for the reason that the most of the hadronic showers develop close to the boundary 
between two modules and hence its influence can be directly quantified and compared. On the other hand, 
the large cone angle will simulate all possible directions of incoming particles and thus can be approximated 
to the calorimeter which covers whole polar angle. The generated data, 10,000 events for smaller angle and 20,000 
events for larger angle for each calorimeter configuration, were subsequently reconstructed and analyzed by a 
standalone program using the org.lcsim framework~\cite{slic}. 

\subsubsection{Digitization}
The Micromegas detector has very fine lateral segmentation (about 1$\times$1\cma), which is read out by 
digital electronics embedded directly on the detector. Calorimeter with digital readout is based on the 
principle that the number of counted hits is proportional to energy deposited in calorimeter. In given case, 
one hit is counted in a 1$\times$1\cma cell only when the energy measured in the cell is higher than a threshold. 
The readout threshold that was used within this study is 0.5 MIP MPV. In order to be as close as possible to the 
real detector conditions, the a full digitization of the signals is performed in the simulation. This includes 
mainly the conversion from energy deposited in 3\mm gas gap from GeV to charge in pC and electronics digitization. 
The most important parameters, such as chamber geometry, mesh transparency, gas amplification, electronics noise, 
and chamber efficiency, where delivered from dedicated laboratory and test beam measurements.  

\begin{center}         
  \begin{figure}[htb]
    \includegraphics[width=0.49\columnwidth]{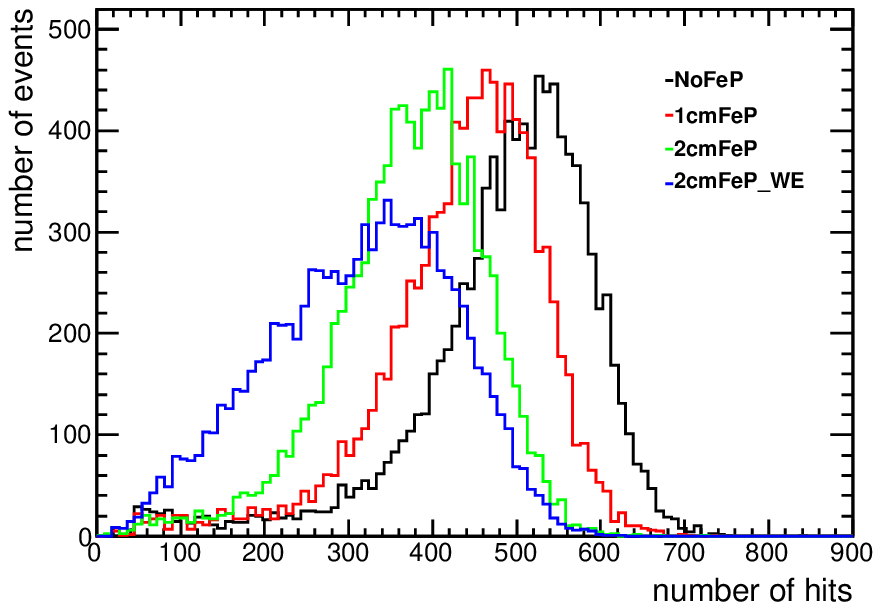} 
    \hfill
    \includegraphics[width=0.49\columnwidth]{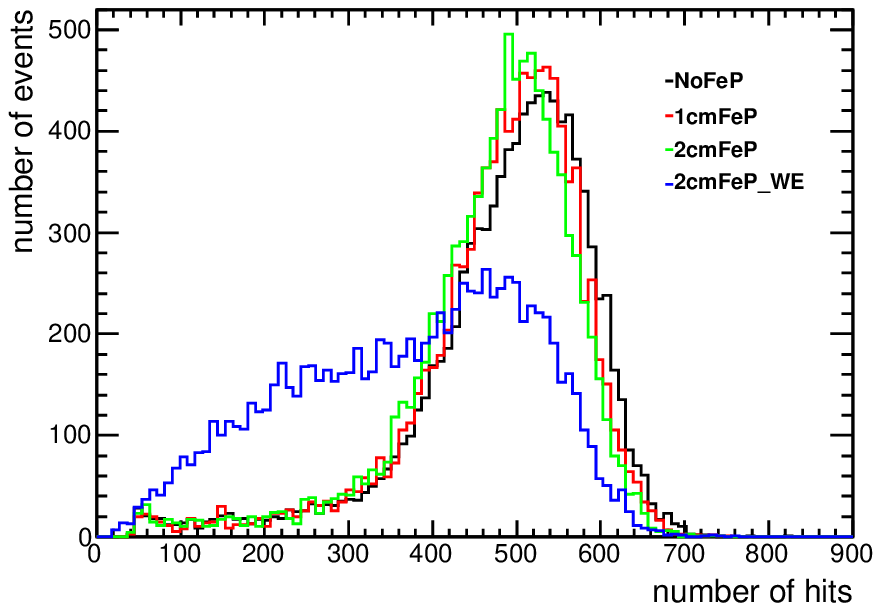}\\ 
    \caption{Distributions of the total number of hits for 50~GeV single pions generated within 2$^\circ$ vertex 
      cone angle for different simulation configurations: ideal geometry ({\it NoFeP}), geometries with 
      various thickness of the supporting plates ({\it 1cmFeP} and {\it 2cmFeP}), and geometry including the 
      electromagnetic calorimeter ({\it 2cmFeP\_WE}). The distributions on the left are for the projective 
      geometry, on the right for the non-projective geometry, respectively.}
    \label{distribution1}
  \end{figure}
\end{center}

\section{Comparison of the projective and non-projective geometry}
\subsection{Calorimeter response}
Figure~\ref{distribution1} shows distributions of the total number of hits counted in calorimeter for 50~GeV single 
pions generated within 2$^\circ$ cone vertex angle for different simulation configurations and for the projective 
and non-projective geometry. In case of the projective geometry, smaller number of hits and thus less visible energy
is seen as the thickness of the absorber plate increases, see Fig.~\ref{distribution1} left. This is because  
the trajectory of the primary pions is close to the boundary and therefor most hadronic showers develop near 
the boundary. Thus the significant part of the shower is absorbed in inactive material of the supporting plate 
and is not measured by the calorimeter. Moreover, in configuration with electromagnetic calorimeter the smallest 
number of hits is counted due to the events with showers that start already inside the electromagnetic calorimeter. 
That is why its distribution has a broad left-hand tail. Of course, this part of the shower energy can be retrieved 
if the electromagnetic calorimeter is active. On the other hand, in case of the non-projective geometry, the number 
of hits collected in calorimeter with and without absorbing plate is almost identical, see Fig.~\ref{distribution1} 
right. This is due to the fact that most primary pions cross the boundary between the modules before starting a
shower, and also because the showers develop in a direction of primary particle. Therefore a very small fraction 
of the showers, if even any, is lost in the supporting plate.         

\begin{center}         
  \begin{figure}[htb]
    \includegraphics[width=0.49\columnwidth]{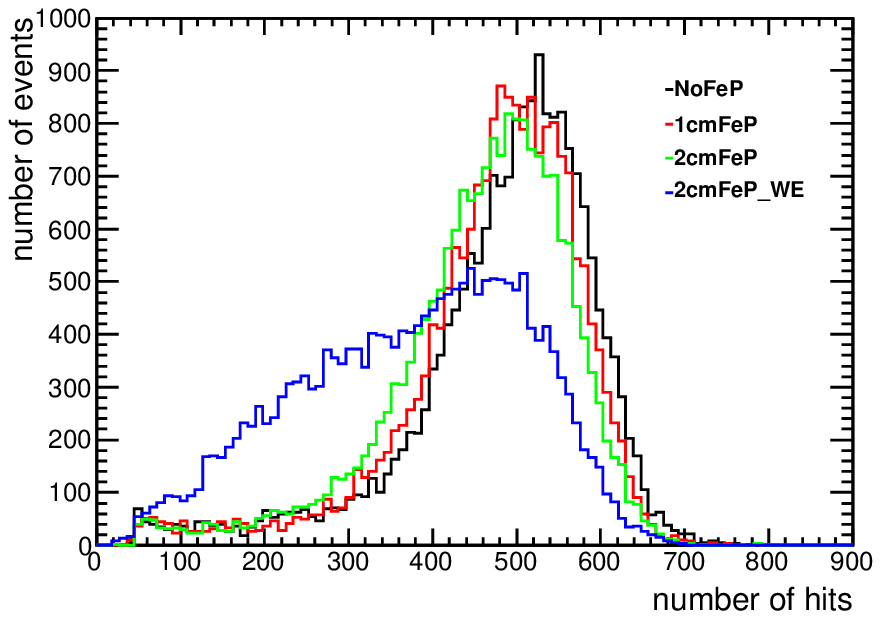}
    \hfill
    \includegraphics[width=0.49\columnwidth]{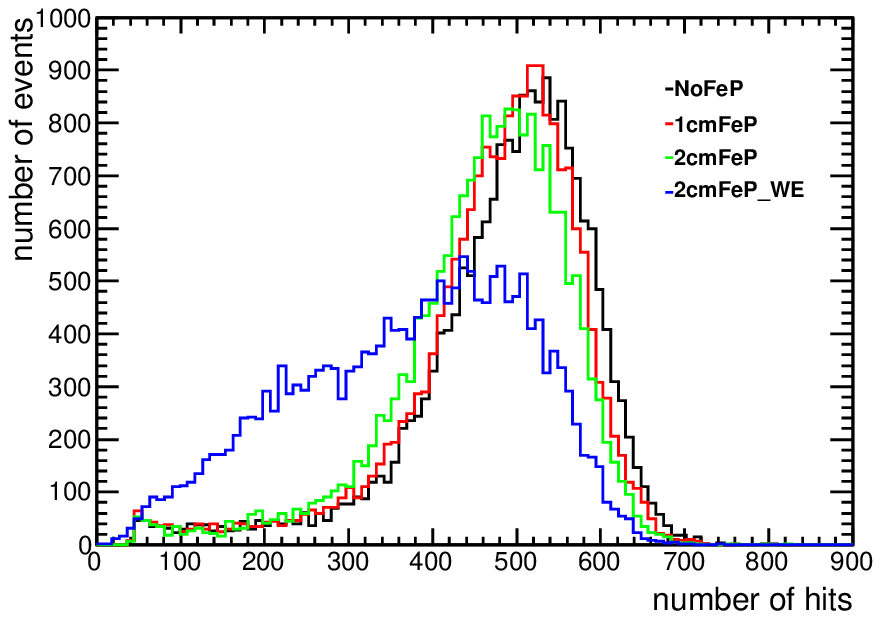}\\
    \caption[]{Similar as the \protect Fig.~\ref{distribution1}, but for 50~GeV single pions generated within 
      15$^\circ$ vertex cone angle. The distributions on the left are for the projective geometry, on the right 
      for the non-projective geometry, respectively.}
    \label{distribution2}
  \end{figure}
\end{center}

Different situation arises in case of the large cone vertex angle (15$^\circ$), for which, both projective 
and non-projective geometry, show similar behaviors, see Fig.~\ref{distribution2}. Though the direction of the 
primary pions is different for both geometries, the total number of the counted hits is almost same. This is 
because the volume of the supporting plate is small with respect to the total volume where energy of the hadronic 
showers can be deposited. Thus the number of hits which is lost due to the crack between the modules is small
in comparison with the average total number of registered hits.

\subsection{Impact of the detector dead zone along the module boundary}
\begin{center}         
  \begin{figure}[htb]
    \includegraphics[width=0.49\columnwidth]{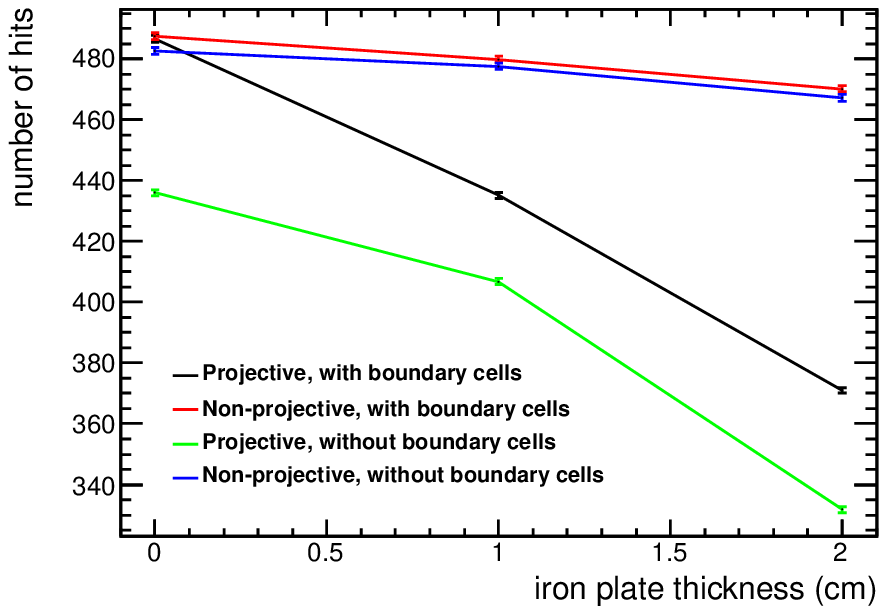}
    \hfill
    \includegraphics[width=0.49\columnwidth]{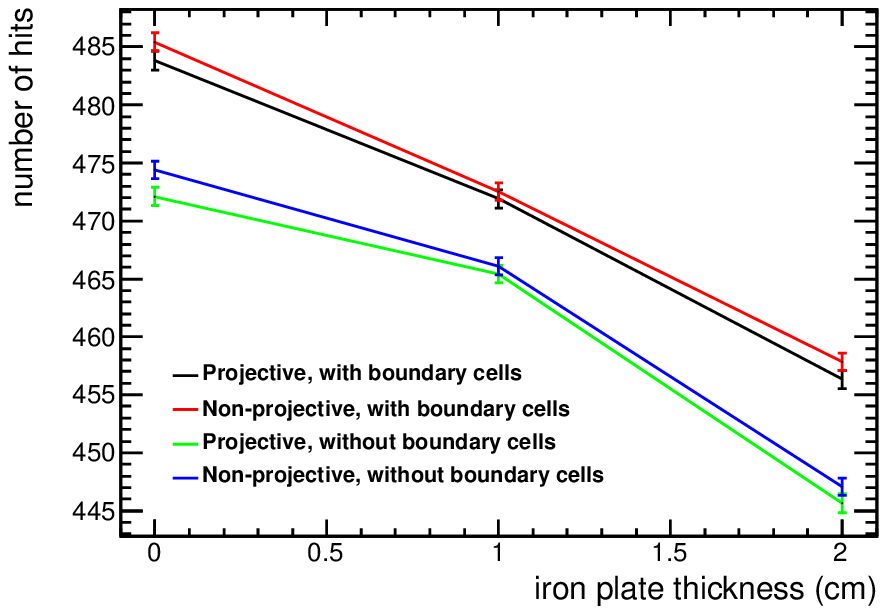}\\ 
    \caption{Mean number of counted hits as a function of the absorbing plate thickness for configuration 
      with and without readout cells along the boundary, and for the projective and non-projective geometry. 
      Results for pions generated within 2$^\circ$ and 15$^\circ$ cone vertex angle are shown in the left and right 
      figure, respectively.}
    \label{meanVsPlate}
  \end{figure}
\end{center}

Due to mechanical constrains, the thickness of the stringers which support the calorimeter modules is limited 
to 2~cm of stainless steel. Thus the thickness of the stringer defines the minimum size of the dead area between
modules. Another contribution to the dead zone is given by the thickness of the frame around the active 
medium, which can be, depending on the technology of the active layer, up to 2~cm. Therefore a total width of
the dead zone is between 2 and 6~cm. Fig.~\ref{meanVsPlate} shows the calorimeter response to single pions measured 
as a mean value of the total number of hits registered in calorimeter for all events versus thickness of the 
supporting plate for the projective and non-projective geometry. The figure compares also results for a full
area active layer and layer with 1~cm dead zone along the edge next to the supporting plate.

\begin{center}         
  \begin{figure}[htb]
    \includegraphics[width=0.49\columnwidth]{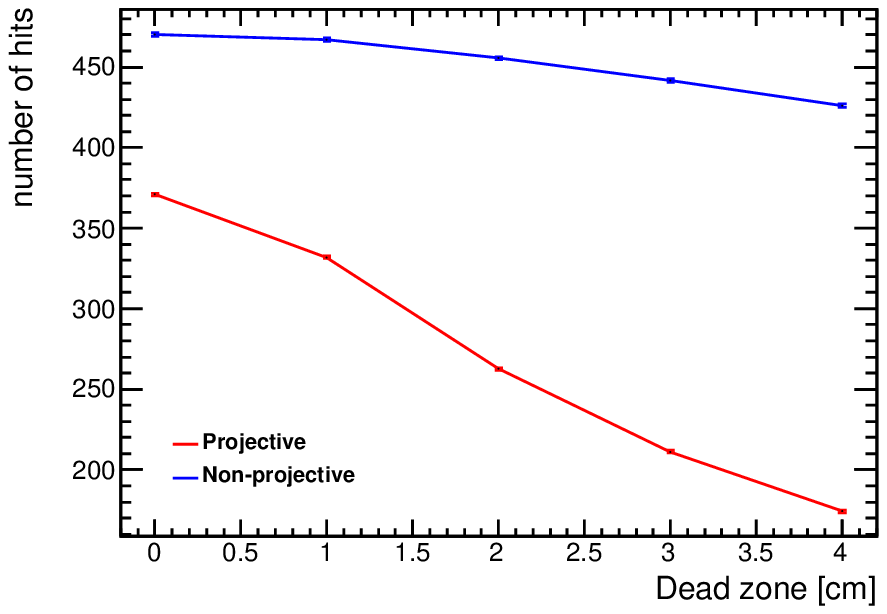}
    \hfill
    \includegraphics[width=0.49\columnwidth]{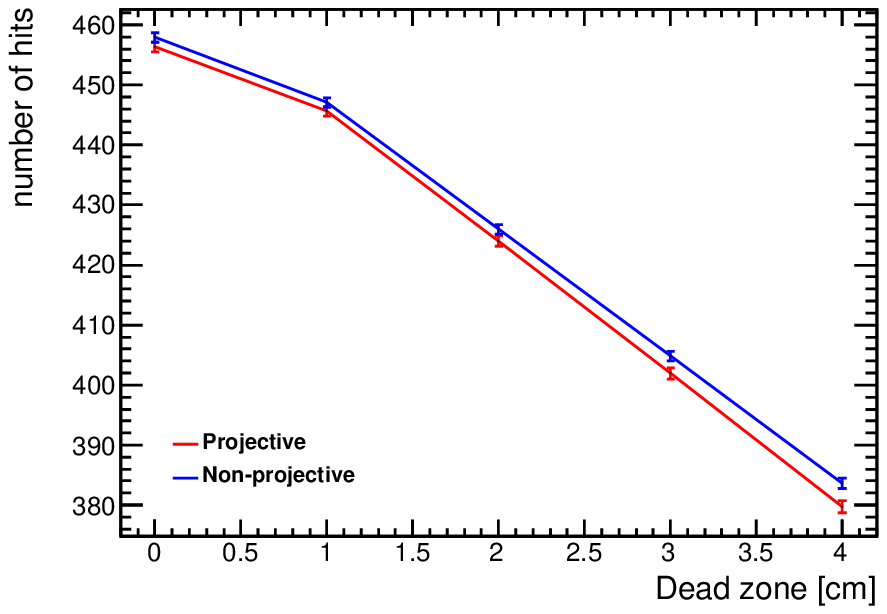}\\ 
    \caption{Mean number of counted hits as a function of the dead zone size for the configuration with 2~cm
      supporting plate and for the projective and non-projective geometry. Results for pions generated within 
      2$^\circ$ and 15$^\circ$ cone vertex angle are shown in the left and right figure, respectively.}
    \label{meanVsDeadZone}
  \end{figure}
\end{center}

Generally, for the reasons explained in previous section, the calorimeter response decrease with increasing 
supporting plate thickness. For small cone vertex angle (see Fig.~\ref{meanVsPlate} left), the projective 
geometry, where showers are father from the boundary, only a small decrease in the number of hits is shown if an 
additional dead zone of 2~cm in the active material (1~cm in each module) is considered. Contrary, the projective 
geometry shows significant decrease in response (about 10\%), because most showers take place close to the boundary. 
For large cone vertex angle (see Fig.~\ref{meanVsPlate} right), same behavior has been found for the projective as 
well as for non-projective geometry with a response variation of -2.9\%/cm of stringers thickness. The response 
falls-off by about 2\% for configuration with 2~cm supporting plate in case of 2~cm dead zone in the active 
layer (1~cm in each module).  

In order to quantify how the size of the dead area in active layer affects the calorimeter response, the size of 
the dead zone has been varied from 0 up to 4 cm for each calorimeter module. Thus the total size including both
the absorber plate and dead areas in active medium have been varied between 2 and 10~cm. As expected, large 
difference have been found between the projective and non-projective geometry for small cone vertex angle, as it is 
shown in Fig.\ref{meanVsDeadZone} left. In case of large vertex cone angle (see Fig.\ref{meanVsDeadZone} right), 
both configurations behave similarly with a response variation of -4.4\%/cm of dead area. 

\begin{center}         
  \begin{figure}[htb]
    \includegraphics[width=0.47\columnwidth]{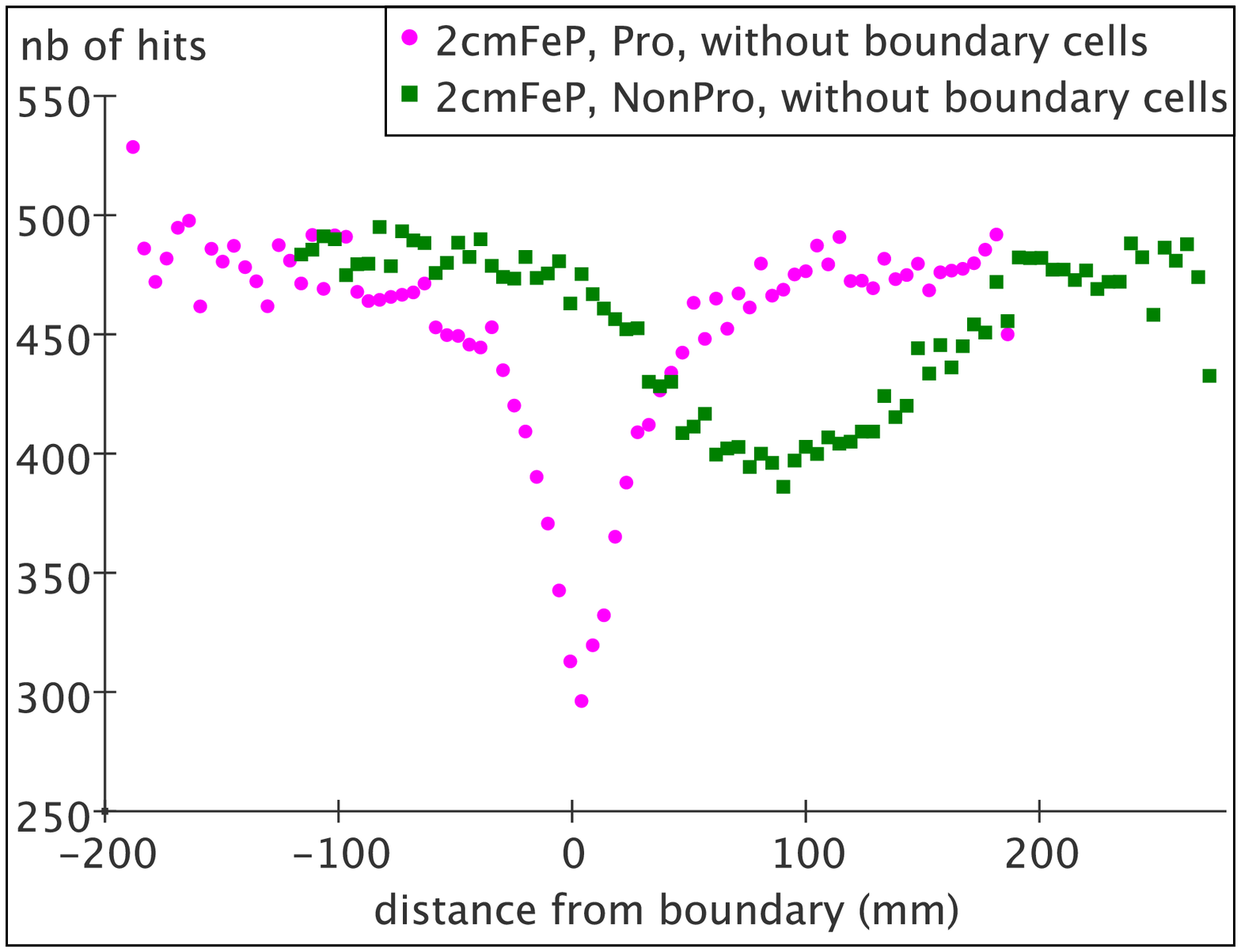}
    \hfill
    \includegraphics[width=0.47\columnwidth]{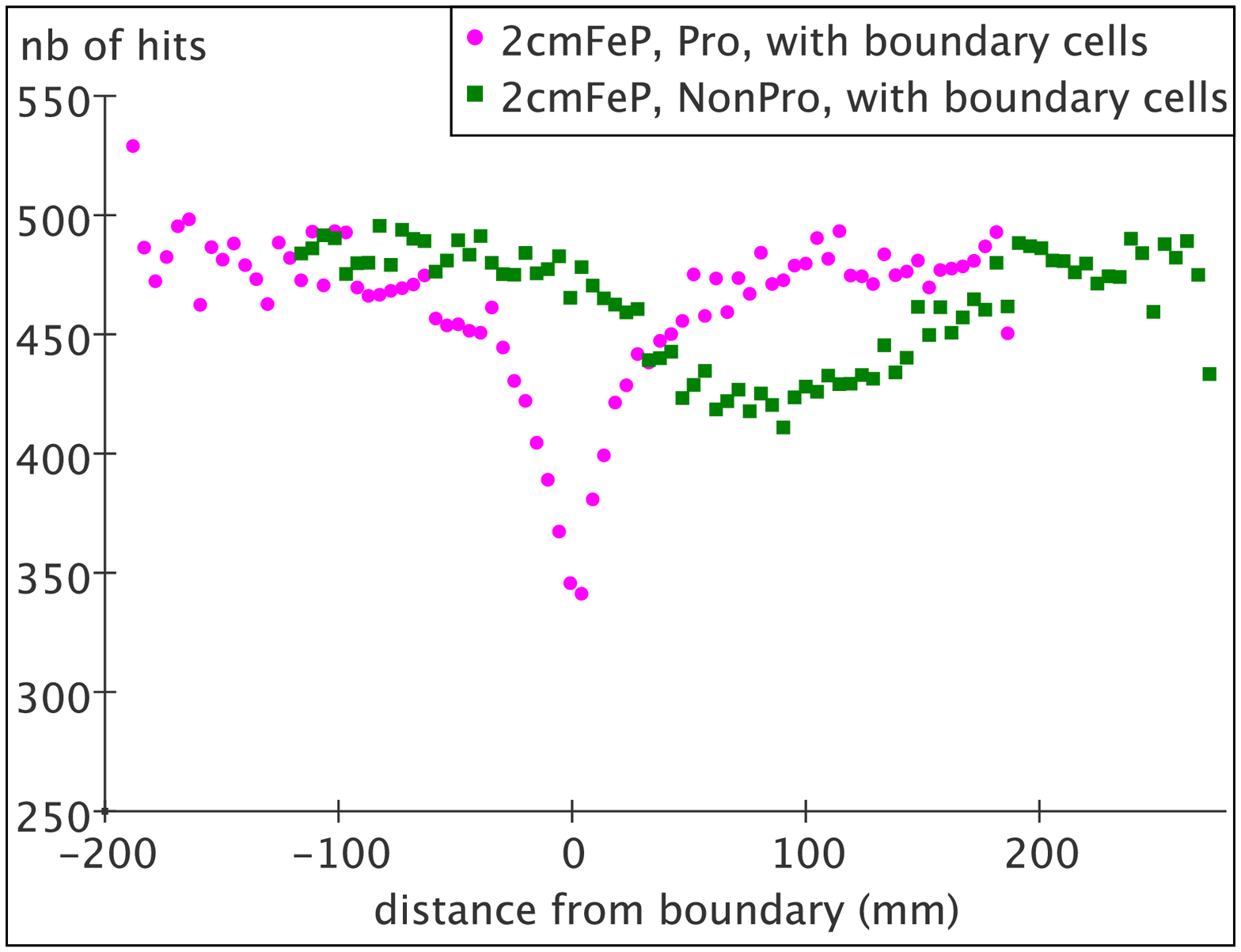}\\ 
    \caption{Mean number of counted hits versus distance from the 2~cm thick supporting plate between modules
      for configuration with (left) and without (right) readout cells along the boundary, and for projective 
      (purple) and non-projective (green) geometry, respectively. The thickness of the supporting plate
      is not included in the distance from boundary (x-axis).}
    \label{aroundCrack}
  \end{figure}
\end{center}

\subsection{Response versus distance from a crack}
An appearance shape of the calorimeter response around a crack for the projective and non-projective geometry 
without readout cells along the boundary (1~cm of dead space along the boundary in each module) is shown in 
Fig.~\ref{aroundCrack} left. The smallest response is observed for the projective geometry, for which the number of 
hits degreases down to 40\% of its maximal value at it is measured far from the crack. As expected, the minimum 
of the response is seen close to the crack. In case of the non-projective geometry, the calorimeter response drops 
by 20\%. Moreover, due to the off-pointing geometry, the minimal value of the response is located about 10~cm away 
from the crack. A similar behavior is expected also in whole non-projective HCAL, where the degrease in response 
due to a crack will be present mainly in trapezoidal modules and smaller part will be seen in rectangular modules.
Finally, if the dead space in the active medium is not considered, the calorimeter response is about 5~\% higher 
than in previous case (see Fig.~\ref{aroundCrack} right).   

\subsection{Comparison for muons}
Another important question in comparing the projective and non-projective geometry is what is the fraction of 
muons that will be not identified in the calorimeter due to cracks. This is an important issue for PFA where the 
total jet energy is computed from individual particle energy contributions in a jet. For performance of the PFA will 
be very useful if muons crossing the calorimeter can be easily identified and assigned to the ones measured in muon 
chambers. 

\begin{center}         
  \begin{figure}[htb]
    \includegraphics[width=0.49\columnwidth]{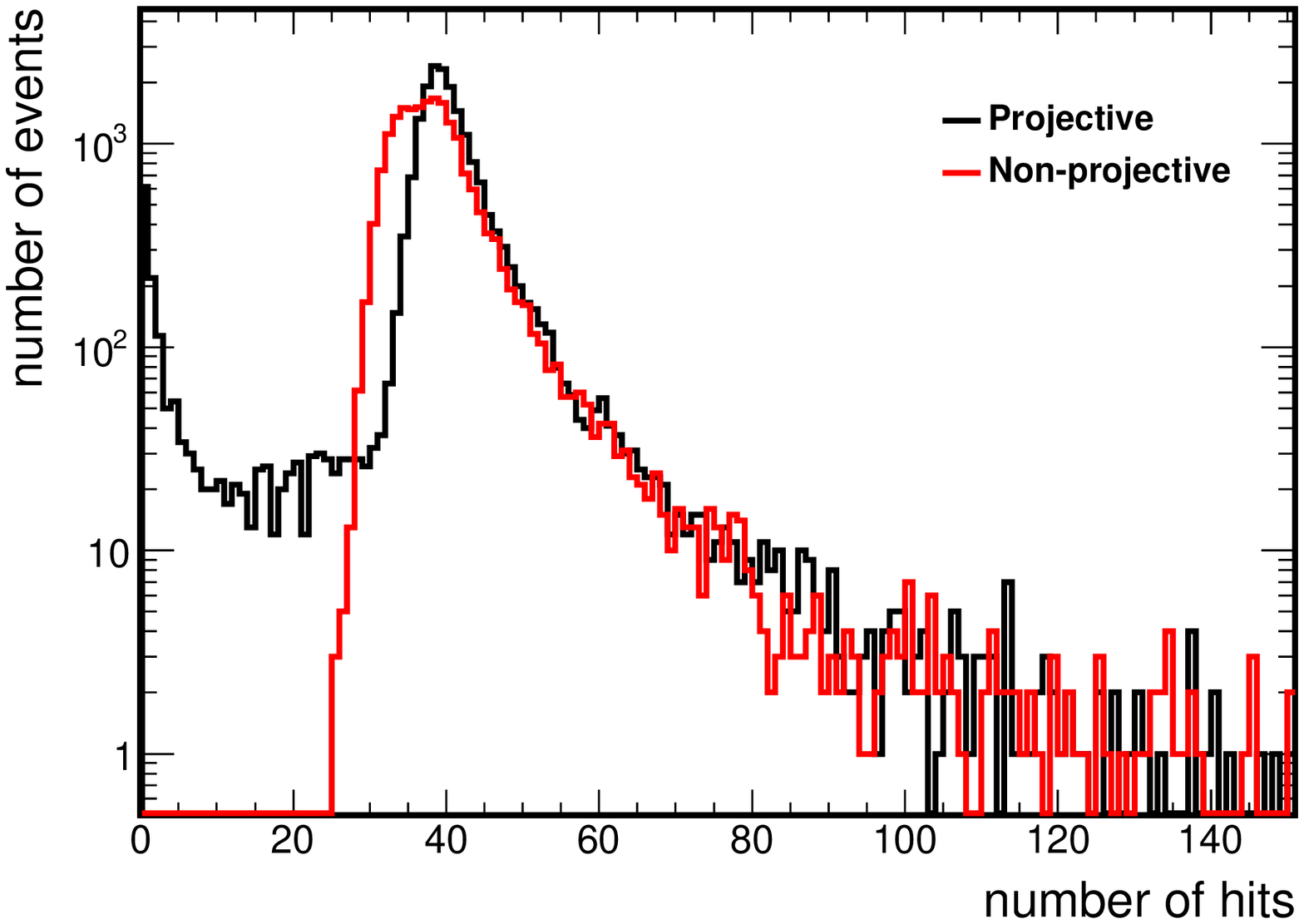}
    \hfill
    \includegraphics[width=0.47\columnwidth]{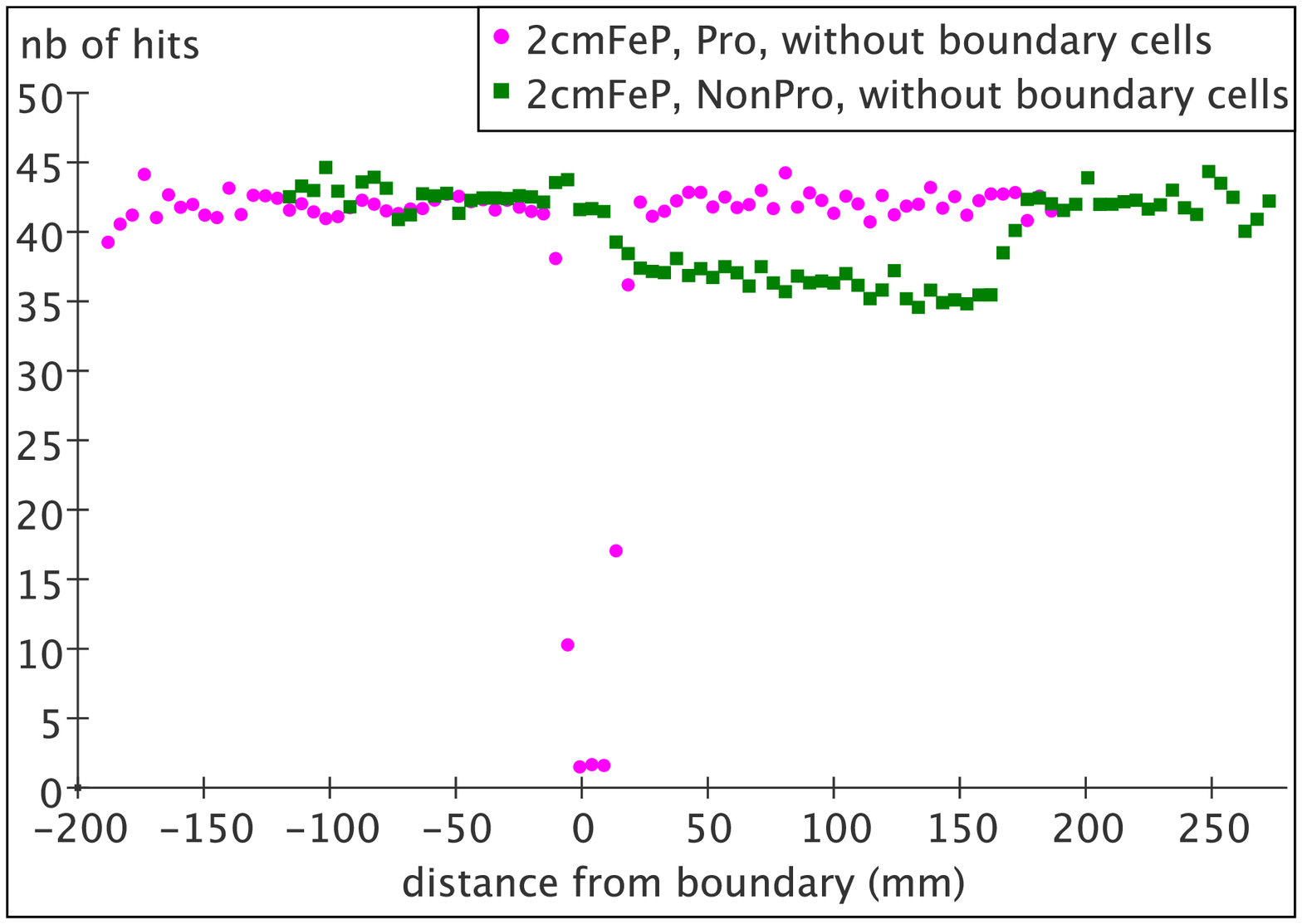}\\ 
    \caption{Distributions of the total number of hits (left) and mean number of hits versus distance from the 2~cm 
      thick supporting plate between modules (right). The distributions are for 50~GeV single muons generated within 
      15$^\circ$ vertex cone angle and for configuration without readout cells along the boundary, and for 
      projective (black or purple) and non-projective (red or green) geometry, respectively.}
    \label{muons}
  \end{figure}
\end{center}

Therefore a test with 50~GeV negative single muons (20,000 events) generated within 15$^\circ$ vertex cone angle has 
been performed and results for the SiD baseline configuration having 2~cm thick supporting plate and 1~cm of a dead 
zone on each module side are shown in Fig.~\ref{muons}. For an ideal muon event, which has passed a calorimeter 
without crossing the module boundary, one hit per layer is measured (this results to 40 hits in SiD HCAL).
In reality, the number of hits can be slightly smaller due to detection inefficiency of the active layer, or 
a bit higher due to interactions along the muon track. If a muon goes through the crack, number of 
registered hits decrease proportionally to the distance which the muon has passed in the crack. In Fig.~\ref{muons} 
left is shown that for the projective geometry, a significant number of muons are not registered at all. On the 
other hand, for every event in non-projective geometry, a clear muon track with more that about 25 hits is recorded.
In other words, if one considers that at least 4 hits are necessary for the muon reconstruction in a calorimeter, 
then about 5~\% of muons events is lost in case of the projective geometry and no one is lost for the non-projective
geometry. It needs to be noted that the fraction of muons, which are not detected due to the crack, will be slightly 
smaller in a real detector due to the presence of the solenoid magnetic field.   

From an appearance shape of the calorimeter response around the crack (Fig.~\ref{muons} right) it is seen that for 
the projective geometry the response is affected in an area of several cm around the crack where the response 
drops sharply to the zero registered hits. In case of the non-projective geometry, the affected area is larger,  
around 20~cm, but the number of registered hits decreases by about 20\%, which still allows to determine a passing
muon close to the module boundary.      

\section{Summary and conclusions}
The projective and non-projective HCAL geometries, which are proposed for SiD detector, have been investigated 
in order to determine the most suitable one. The comparison for various design configurations allowed to evaluate 
the impact of the supporting stringers and dead areas along the module boundary which has been studied with single 
pions and muons. The impact of the boundary between modules is clearly seen for area about 20~cm around a crack. 
This corresponds to a cone angle of 8$^\circ$ which means that for 12 modules the affected area is about 26\% of the 
whole polar angle. Thus, it can be concluded that the decrease of the calorimeter response close to the boundary 
(for small angles of impinging particles) is significantly smaller for the non-projective geometry. If the 
impinging particles are distributed in the large polar angles, the global response is similar for both geometries. 
In this case a response variation of -2.9\%/cm of stringers thickness and -4.4\%/cm of dead area have been found. 
This confirms that for base-line SiD design (2 cm thick stringers and 1~cm of dead area in each module) an average 
response will decrease by about 7 to 8\% around a crack. The advantage of the non-projective geometry is that no 
muons will be lost in a crack, contrary to about 5\% of muon events which will be not identified in the projective 
geometry. On the other hand, the drawback of the non-projective geometry is the mechanical design requiring two 
different module shapes in comparison with the projective geometry where all the modules are identical. All this 
aspects need to be carefully weighted in order to give a final decision. 

Finally, it needs to be pointed out that because this study has been performed for the simplified geometry 
description with single pions and muons, it provides very important, but only the first image how the cracks may 
impact on the calorimeter performance. Also the absence of the magnetic field and information from the other detectors 
subsystems need to be taken into account for the reconstruction performance in a real detector. Therefore, it is advisable to 
perform a complementary study with whole SiD detector and real physics analysis of events with jets.      

\acknowledgments
We would like to thank to our colleague A.~Espargili\`{e}re for implementing the Micromegas digitization driver that 
has been used within this study. Also we would like to express our thanks to the SLIC and lcsim developers whose
software has been used and tested in this study and who have always promptly answered all our technical questions 
concerning this simulation and analysis framework.

\clearpage
\appendix
\section{Detailed description of the geometry as it is used in simulation}
\begin{center}         
  \begin{figure}[htb]
    \includegraphics[width=0.96\columnwidth]{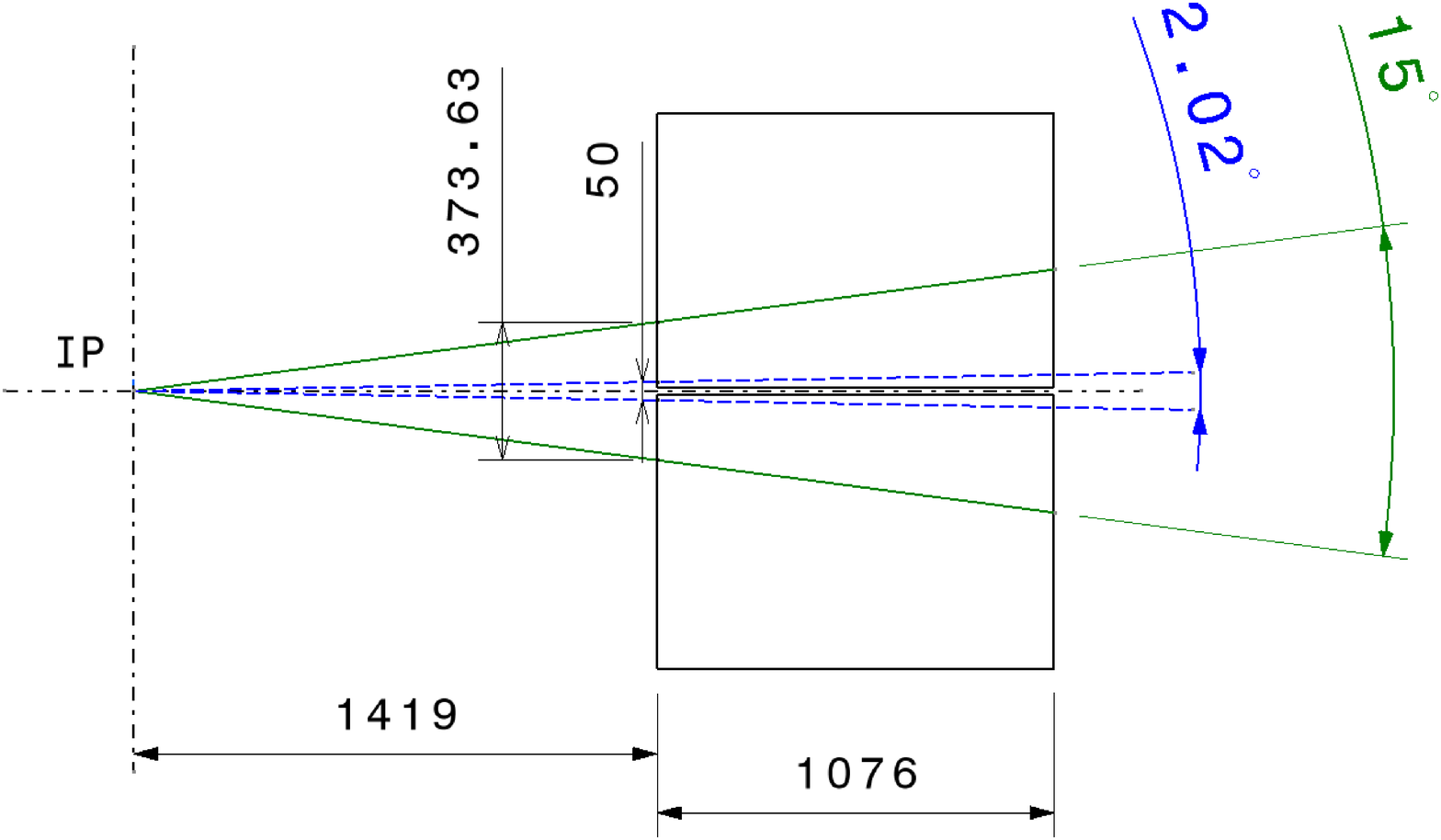}\\
    \vfill
    \includegraphics[width=0.96\columnwidth]{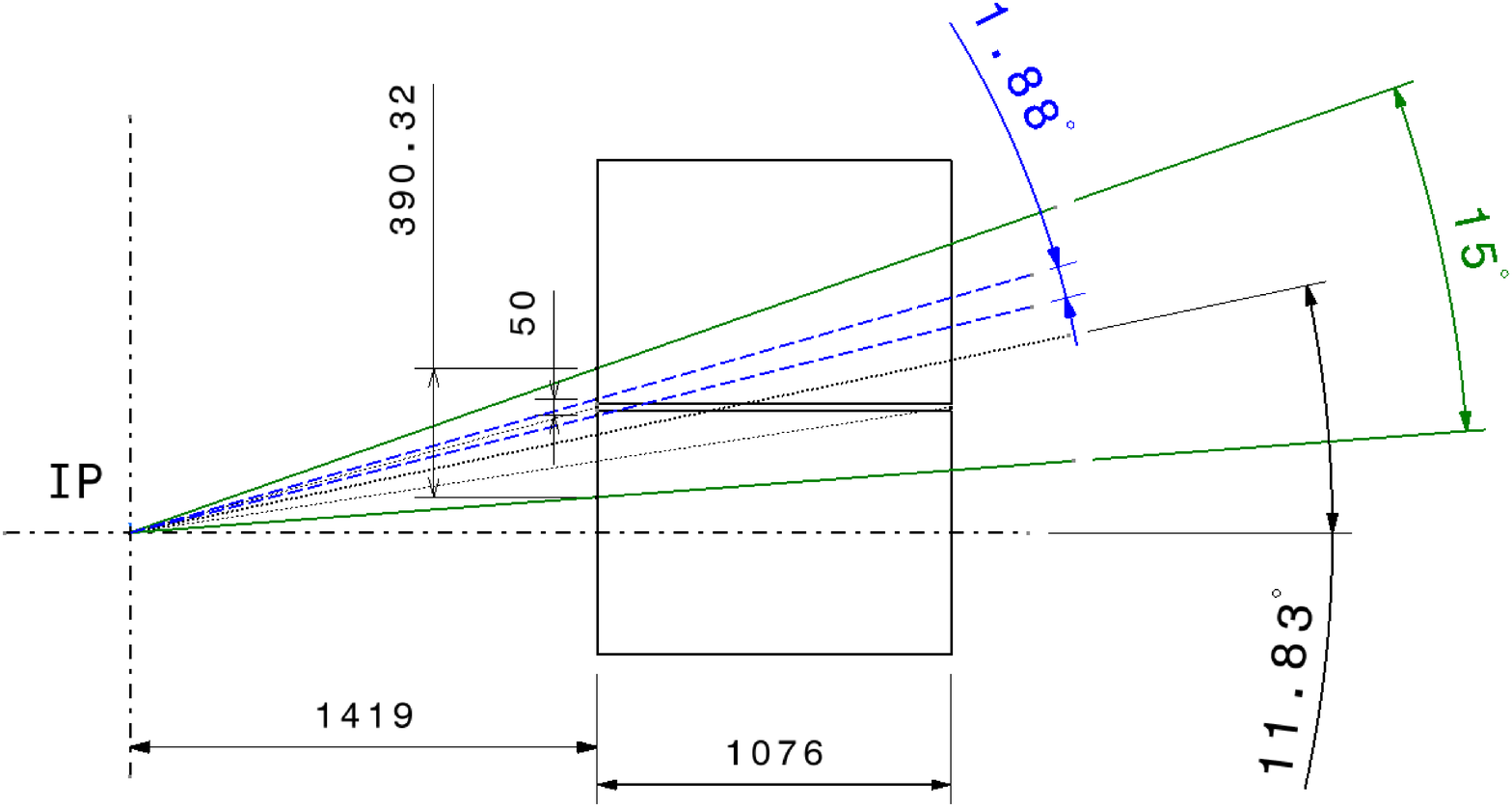}\\ 
    \caption{Description of the projective (top) and non-projective (bottom) geometry as it was 
      used in simulation. In blue and red colors are shown cross-sections of the cones defining 
      direction of the particle coming from the interaction point (IP) and impinging on the front 
      face of the calorimeter. For each cone the cone angle and impact area is shown.}
    \label{appxGeometry}
  \end{figure}
\end{center}
\end{document}